\documentstyle[aps,epsf,multicol,amsmath,amsfonts,graphicx,bbm]{revtex}

\begin{document}

\title{Secure direct communication using entanglement}

\author{Kim Bostroem}
\address{Institut f\"ur Physik, Universit\"at Potsdam, 
14469 Potsdam, Germany
}
\date{\today}
\maketitle

\begin{abstract}

A novel communication protocol based on an entangled pair of qubits is presented, allowing secure direct communication from one party to another without the need for a shared secret key. Since the information is transferred in a deterministic manner, no qubits have to be discarded and every qubit carries message information. The security of the transfer against active and passive eavesdropping attacks is provided. The detection rate of active attacks is at least 25\%. The protocol works with a quantum efficiency of 1 bit per qubit transmitted.

\end{abstract}

\begin{multicols}{2}
\narrowtext

\section{Introduction}

What is secure direct communication? Traditionally, secure communication schemes based on quantum mechanics are \emph{non-deterministic} \cite{BB84,Ekert,Bruss,Bourennane,Cerf}: Alice, the sender, cannot determine which bit value Bob receives through the secure quantum channel. Such non-deterministic communication can be used to establish a \emph{secret key} between Alice and Bob. Whenever an eavesdropper tries to extract information from the quantum channel, he influences the transmitted state and can be detected with some probability. If Alice and Bob are virtually sure that a certain random subsequence of bits has been transmitted secretely, Alice can use the remaining subsequence as a shared secret key to encrypt her message, send the encrypted message to Bob through a non-secret channel and then Bob uses the shared key to decrypt the message. It is a common belief that every secure quantum communication protocol should work that way. 

Recently, however, a \emph{deterministic} quantum cryptographic protocol has been presented \cite{Almut1,Almut2}, which I will refer to as the \emph{BEKW protocol}. Against the paradigm of quantum cryptography, the information is sent \emph{directly} from Alice to Bob. Alice uses a secret key to encrypt her message before sending it through a quantum channel. If she is virtually sure that no eavesdropper was in the line, Alice publishes the secret key so Bob can read the message. This is a different concept of quantum cryptography, and I will refer to it as \emph{secure direct communication} as opposed to \emph{quantum key distribution}. 
In the present paper, another deterministic cryptographic scheme is presented which has significant advantages against other schemes:
\begin{enumerate}
\item
High quantum efficiency: 1 bit per qubit.
\item
Reliable security: Detection probability $d=1/4$ per control qubit.
\item
Deterministic: No qubits are discarded. The message is send directly.
\end{enumerate}
The magic ingredient for the protocol is entanglement. Alice has to prepare and measure Bell pairs, and she must be able to store one of the entangled qubits for a certain time (some milliseconds will do).
Even if it might still be difficult to prepare and to measure the Bell states, and in particular to store entangled qubits, the presented scheme contains a completely new approach to assure the security of a communication.
In order to illustrate the quality of the protocol, let us compare it with other schemes. In non-deterministic protocols, Alice and Bob choose at random one of several possible bases to prepare and/or to measure the transmitted qubit. If their choices does not coincide, the qubit is discarded since the outcome is completely uncorrelated. In the BB84 protocol \cite{BB84} every second qubit is discarded that way, so the quantum transmission rate,
\begin{equation}
	R_q=\frac{\text{N(usable bits)}}
	{\text{N(qubits)}},
\end{equation} 
is $R_q=0.5$. The protocol needs an additional classical channel carrying 2 bits per transmission (the choices of basis), so the total rate
\begin{equation}
	R_{\rm tot}=\frac{\text{N(usable bits)}}
	{\text{N(qubits)}+\text{N(bits)}},	
\end{equation}
is $R_{\rm tot}=0.5/(1+2)=1/6$. Since the usable bits form the shared secret key, there is still the need for a subsequent classical transfer of the same bit size carrying the message information.
The same goes for the Ekert scheme \cite{Ekert}. In the BEKW protocol, every transmitted qubit can directly be used for the message, but only two of them carry one message bit, so $R_q=0.5$. The additional classical channel carries 2 bits per transmission, so $R_{\rm tot}=1/6$.
In the protocol presented here, two qubits and 2 bits of message information are transmitted, hence $R_q=1$. The classical channel transfers 2 bits, so $R_{\rm tot}=2/(2+2)=1/2$. No further transmission is required.
All the rates given above do not include \emph{control bits}. So let us roughly compare the security of the protocols, which depends on the detection probability $d$ per transmitted control qubit. 
The BB84 and the Ekert protocol provide $d=1/4$, the BEKW protocol provides $d=1/6$, and the protocol presented here provides $d=1/4$. 

\section{Double dense coding}

Let us start with a protocol that is not designed for security but for highly efficient use of quantum resources. Afterwards we will construct from there the secure protocol. 

\begin{figure}
\begin{center}
	{\includegraphics[width=0.25\textwidth]{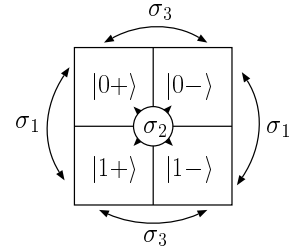}}
	\caption{\small Transformation table.}\label{trafotable}
\end{center}
\end{figure}

\begin{figure}
\begin{center}
	{\includegraphics[width=0.3\textwidth]{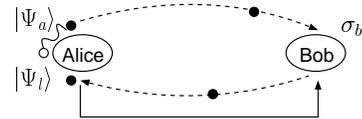}}
	\caption{\small Double dense coding: 4 bits are exchanged by one entangled qubit travelling forth and back.}\label{ab}
\end{center}
\end{figure}

Alice controls a device that is able to prepare two qubits in one of the Bell states
\begin{eqnarray}
	|0\pm\rangle&=&\frac1{\sqrt2}(|00\rangle\pm|11\rangle),\\
	|1\pm\rangle&=&\frac1{\sqrt2}(|01\rangle\pm|10\rangle),
\end{eqnarray} 
forming the orthogonal Bell basis ${\cal B}_{\rm Bell}$. 
Let us bring the Bell states in some order,
\begin{eqnarray}
	|\Psi_0\rangle=|0+\rangle,\quad|\Psi_1\rangle=|0-\rangle,\\
	|\Psi_2\rangle=|1+\rangle,\quad|\Psi_3\rangle=|1-\rangle.
\end{eqnarray}
Alice decides to send the value $a\in\{0,1,2,3\}$ by preparing the \emph{initial state} $|\Psi_a\rangle\in{\cal B}_{\rm Bell}$, which she can easily look up in Fig.~\ref{alicescheme}.
\begin{figure}
\begin{center}
	{\includegraphics[width=0.2\textwidth]{alicescheme.eps}}
	\caption{\small Alice's coding scheme.}\label{alicescheme}
	\end{center}
\end{figure}
After the state is prepared, Alice sends only the \emph{second} qubit, which we call the \emph{travel qubit}, and she keeps the first one, which we call the \emph{home qubit}. 
Bob receives the travel qubit and performs one of the four unitary operations
\begin{eqnarray}
	\sigma_0&=&{\mathbbm1}~=|0\rangle\langle0|+|1\rangle\langle1|,\\
	\sigma_1&=&\sigma_x=|1\rangle\langle0|+|0\rangle\langle1|,\\
	\sigma_2&=&\sigma_y=i|1\rangle\langle0|-i|0\rangle\langle1|,\\
	\sigma_3&=&\sigma_z=|0\rangle\langle0|-|1\rangle\langle1|,
\end{eqnarray}
forming the set ${\cal P}$.
These matrices are unitary and hermitian at the same time, so each one is its own inverse, $\sigma_i^2={\mathbbm1}$,
and although they are \emph{local} operations, they switch between the \emph{non-local} Bell states in the following manner:
\begin{eqnarray}
	({\mathbbm1}\otimes\sigma_0)|n\pm\rangle&\sim&|n\pm\rangle,\\
	({\mathbbm1}\otimes\sigma_1)|n\pm\rangle&\sim&|m\pm\rangle,\\
	({\mathbbm1}\otimes\sigma_2)|n\pm\rangle&\sim&|m\mp\rangle,\\
	({\mathbbm1}\otimes\sigma_3)|n\pm\rangle&\sim&|n\mp\rangle,
\end{eqnarray}
with $n\in\{0,1\}$ and $m=1-n$. 
For the sake of simplicity, we have neglected the irrelevant global phase on the righthand side.
Bob receives the travel qubit from Alice and encodes $b\in\{0,1,2,3\}$ by performing $\sigma_b$ on it. After that, he sends the travel qubit back to Alice. 
Alice then performs a Bell measurement on both qubits resulting in one of the four Bell states.
Let us call this resulting state the \emph{final state} $|\Psi_f\rangle\in{\cal B}_{\rm Bell}$.
Since she knows the state $|\Psi_a\rangle$ she initially prepared, she now also knows the operation $\sigma_b$ Bob has performed by solving
\begin{eqnarray}\label{alicecalc}
	 |\Psi_f\rangle&\sim&({\mathbbm 1}\otimes\sigma_b)|\Psi_a\rangle.
\end{eqnarray}
Since there are 4 possible operations \emph{intentionally} performed by Bob, Alice has received 2 bits of \emph{deterministic information} from him. If she now tells Bob via a classical channel the final Bell state $|\Psi_f\rangle$ resulting from her measurement, then Bob can calculate the initial state $|\Psi_a\rangle$ by inverting~(\ref{alicecalc}).
Since there are 4 possible initial Bell states that Alice intentionally prepared, Bob has received 2 bits of deterministic information from her.
Instead calculating, Alice and Bob can also look at the transformation table in Fig.~\ref{trafotable} to decode the other one's message.
The table indicates the effect of the local operations $\sigma_b$ to the Bell state. For example, $\sigma_2$ transforms $|0+\rangle\leftrightarrow|1-\rangle$ and $|0-\rangle\leftrightarrow|1+\rangle$. Not indicated is the effect of $\sigma_0={\mathbbm1}$, which does nothing. The transmission scheme is depicted in Fig.~\ref{ab}.
The efficiency of this protocol is high: 1 qubit travels forth and back between Alice and Bob and 4 bits of information are exchanged, which makes a quantum transmission rate of 2 bits per transmitted qubit. The origin of such high efficiency is the same as in the \emph{dense coding} protocol \cite{densecoding}: entanglement. The present protocol works in a similiar manner and since it works in two directions simultaneously, let us call it a \emph{double dense coding} protocol.

\section{Eavesdropping}

Say there is eavesdropper Eve having full access to the quantum channels between Alice and Bob. Whatever measurement Eve might be performing, the travel qubit carries no information about the initial state prepared by Alice, because the state of the travel qubit equals for \emph{any} initial Bell state the complete mixture,
\begin{eqnarray}
	\forall a:\quad \rho_1&=& {\rm Tr}_2\{|\Psi_a\rangle\langle\Psi_a|\}
	=\frac12{\mathbbm1}.
\end{eqnarray}
However, Eve can gain some information about \emph{Bob's} message. If Eve measures the travel qubit coming from Alice in some basis, resends it in a certain state, and then measures the state of the qubit returned by Bob, she can gain 1 bit of information about the action that was performed on the qubit. Such an attack is called \emph{intercept-resend attack}.

\begin{figure}
\begin{center}
	{\includegraphics[width=0.28\textwidth]{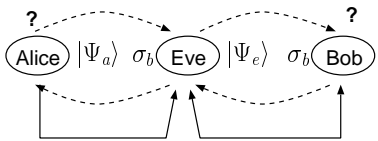}}
	\caption{\small Eve as the ``man in the middle'': total control.}\label{aebcapture}
	\end{center}
\end{figure}

But Eve is \emph{really} smart. She captures the qubit as it travels from Alice to Bob. Then she prepares another ``evil'' Bell state $|\Psi_e\rangle$, sends one of its qubits as a travel qubit to Bob, who performs his operation and sends it back to Alice. But Eve also catches that qubit and performs a Bell measurement on both evil qubits.
Now she knows exactly the operation $\sigma_b$ that Bob has performed and hence got his message. She applies the operation $\sigma_b$ to the captured ``friendly'' travel qubit and sends it back to Alice, who performs a Bell measurement resulting in the final state $|\Psi_f\rangle$. When Alice announces the final state $|\Psi_f\rangle$, Eve can calculate the initial state $|\Psi_a\rangle$ and hence got Alice's message.
Bob performs the same calculation and also obtains $|\Psi_a\rangle$. So even if Alice and Bob would sacrifice their bits to compare them publicly, they would not notice that Eve was in the line.
To Alice Eve acts like Bob and to Bob Eve acts like Alice.
Such an evil thing is called a \emph{man-in-the-middle attack} and is depicted in Fig.~\ref{aebcapture}.
There is no protection against a man-in-the-middle attack except: a \emph{public channel}. This is a channel that spreads its information content all over the world. Eve can read and write to the channel, but she cannot \emph{manipulate} it. If even the public channel is under full control of Eve, then let us call this a \emph{strong} man-in-the-middle attack or a \emph{Berlusconi attack}. The protocol presented here is secure against \emph{weak} man-in-the-middle attacks: it does not allow Eve to manipulate the public channel. A fairly good public channel is already given by an ordinary radio signal.

Apart from \emph{active attacks} described above, there can also be \emph{passive attacks}. Once the final state $|\Psi_f\rangle$ is published, the messages of Alice and Bob are strongly correlated via~(\ref{alicecalc}). An eavesdropper could try a so-called \emph{known plaintext attack}: If there is some part of Alice's or Bob's message that is likely to appear in the message at a certain position (e.g. ``Hi Baby'' or ``See you''), the eavesdropper can \emph{immediately} decypher the other one's part of the message at the same position. 
No serious cryptosystem should allow that.

\section{The final protocol}

First we abandon the ``full duplex'' property of the double dense coding scheme and allow only Alice to send messages to Bob. 
Bob uses a random number generator to choose the operation $\sigma_b$ he performs on the travel qubit. Bob's random sequence of operations acts like a secret key on Alice's message.

Next we have to make sure that Eve is detected with nonzero probability while she is trying to do active attacks. 
Let us introduce two modes, the \emph{message mode} and the \emph{control mode}.
In message mode, Alice and Bob perform the double dense coding protocol, but Alice does not tell the final state she has measured.
With probability $\lambda_c$ Bob switches from message mode to control mode.
Now, instead of performing his operation, Bob randomly chooses a basis ${\cal B}_i$ out of the two bases ${\cal B}_0=\{|0_0\rangle,|1_0\rangle\}$ or ${\cal B}_1=\{|0_1\rangle,|1_1\rangle\}$, where
\begin{eqnarray}
	|0_0\rangle&=&|0\rangle,~~~~~~~~~~~~~~~~~
	|1_0\rangle=|1\rangle,\\
	|0_1\rangle&=&\frac1{\sqrt2}(|0\rangle+|1\rangle),~~
	|1_1\rangle=\frac1{\sqrt2}(|0\rangle-|1\rangle).
\end{eqnarray}
He performs a \emph{measurement} in the basis ${\cal B}_i$ on the travel qubit and instead returning Alice the travel qubit, Bob sends her his choice of basis and his measurement result through the public channel.
Now Alice also switches to control mode and performs a measurement in the same basis on her home qubit. Then she looks if the result is correlated or anticorrelated to Bob's result. Since she (and only she) knows the initial state $|\psi_a\rangle$, she sees if the correlation is wrong or right. If the correlation is wrong, she knows that someone was in the line and stops the communication. 
If the correlation is correct, she repeats the preceding run in message mode because the last qubit she sent has been sacrificed.

At the end of an undisturbed communication sequence, Alice sends Bob the list of final Bell states she has measured. Since Bob knows the sequence of operations he performed, he can decypher Alice's messages by solving~(\ref{alicecalc}) or using Fig.~\ref{trafotable}.

Now here is the explicit algorithm realizing the protocol. 
Alice wants to communicate the message $\boldsymbol a=a_1\cdots a_N$, where $a_n\in\{0,1,2,3\}$.
\begin{enumerate}
\renewcommand{\labelenumi}{p.\arabic{enumi})}
\setcounter{enumi}{-1}
\item\label{pinit}
Protocol is initialized. $n=0$, $\boldsymbol f=\boldsymbol b=\emptyset$.
\item\label{pstart}
$n=n+1$.
Alice prepares two qubits in the Bell state $|\psi_{a_n}\rangle$.
\item\label{pstart2}
She keeps the first qubit, the \emph{home qubit}, and sends the other one, the \emph{travel qubit}, to Bob.
\item
Bob receives the travel qubit. With probability $\lambda_c$ he switches to control mode and proceeds with~c.1, else he proceeds with m.1.
\begin{enumerate}\renewcommand{\labelenumii}{c.\arabic{enumii})}
	\item\label{cstart}
	Bob chooses at random a basis ${\cal B}_i\in\{{\cal B}_0,{\cal B}_1\}$.
	\item
	He measures the travel qubit in the basis ${\cal B}_i$ and obtains the value $j\in\{0,1\}$ with equal probability.
	\item
	He sends $ij$ through the public channel to Alice.
	\item
	Alice receives $ij$ through the public channel, switches to control mode and measures her home qubit in the basis ${\cal B}_i$ resulting in the value $k$.
	\item
	$(|\psi_{a_n}\rangle=|0\pm\rangle\wedge j\neq k)\vee
	(|\psi_{a_n}\rangle=|1\pm\rangle\wedge j=k)$: Eve is detected. Abort or Goto p.\ref{pinit}. \\
	Else set $n=n-1$ and goto~p.\ref{pstart}.
\end{enumerate}
\begin{enumerate}\renewcommand{\labelenumii}{m.\arabic{enumii})}
	\item\label{mstart}
	Bob takes a number $b_n\in\{0,1,2,3\}$ from his random number generator, appends the value $b_n$ to the list $\boldsymbol b$, applies the operation $\sigma_{b_n}$ to the travel qubit and
	sends it back to Alice.
	\item
	Alice receives the travel qubit and makes a Bell measurement on both qubits resulting in the \emph{final state} $|\psi_{f_n}\rangle\in{\cal B}_{\rm Bell}$. She appends the value $f_n$ to the list $\boldsymbol f$.
	\item
	$n<N$: Goto~p.\ref{pstart}.\\
	$n=N$: Goto~p.\ref{pfinal}.
\end{enumerate}
\item\label{pfinal}
Alice sends the list $\boldsymbol f$ to Bob.
\item
For each $(f_n,b_n)$ Bob decodes $a_n$ via
\begin{eqnarray}\label{bobcalc}
	 |\Psi_{a_n}\rangle&\sim&({\mathbbm 1}\otimes\sigma_{b_n})|\Psi_{f_n}\rangle,
\end{eqnarray}
or by looking at Fig.~\ref{trafotable}.
\end{enumerate}
The published sequence $\boldsymbol f$ is completely uncorrelated with Alice's message $\boldsymbol a$, since the sequence $\boldsymbol b$ of operations is a random sequence. Just like a \emph{one-time pad} scheme, there is no way to break the cryptosystem by passive attacks, if the sequence $\boldsymbol b$ is truly random and used only once, which we assume here.
After the qubit has arrived at Bob, with probability $\lambda_c$ he activates control mode, so Eve has no chance to adapt her strategy. If she has been replacing the travel qubit with another qubit, using \emph{intercept-resend} or \emph{man-in-the-middle} strategy, it is not entangled with Alice's home qubit, and with probability $1/2$ she has forwarded a state in a basis different from Bob's. If so, then with another probability of $1/2$ the measurements of Alice and Bob, which are performed in the same basis, show the wrong type of correlation, hence with total probability $d=1/4$ Eve is detected. 
The public channel is needed to synchronize Alice and Bob in control mode.
\begin{figure}
\begin{center}
	{\includegraphics[width=0.26\textwidth]{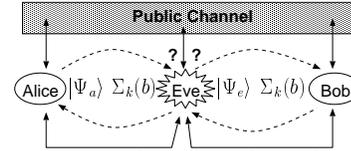}}
	\caption{\small Quantum correlations and a public channel unveil evil Eve.}\label{aebpublic}
	\end{center}
\end{figure}
The probability for a control transfer is $\lambda_c$, so for $N$ protocol runs the number of control runs is $N_c=\lambda_c N$, and with Eve attacking all the time, the probability that she stays undetected reads
\begin{equation}
	\overline D(N)=(1-d)^{\lambda_c N}
	=\left(\frac34\right)^{\lambda_c N}.
\end{equation}
The above value can be made arbitrarily small by choosing an $\lambda_c$ and $N$ appropriately. 
Concluding, the protocol is asymptotically secure against active and passive attacks.

\section{Acknowledgements}

I had exciting and enlightening discussions with Timo Felbinger, Almut Beige, Luke Rallan, Jens Eisert, Martin Plenio, Sougato Bose, and others. This work is supported by the Deutsche Forschungsgemeinschaft (DFG).


\end{multicols}
\end{document}